\documentclass[runningheads]{llncs}

\usepackage[T1]{fontenc}
\usepackage{amsmath,amssymb}
\usepackage{graphicx}
\usepackage{booktabs}
\usepackage{array}
\usepackage{listings}
\usepackage{xspace}
\usepackage{tikz}
\usetikzlibrary{arrows.meta,positioning,fit,backgrounds,calc,shapes.geometric}
\usepackage{pgfplots}
\pgfplotsset{compat=1.17}
\usepackage{colortbl}
\definecolor{rowtint}{gray}{0.92}
\usepackage[hidelinks]{hyperref}
\usepackage{microtype}
\newcommand{\fab}{\textsc{RV-Fabric}\xspace}
\newcolumntype{R}[1]{>{\raggedright\arraybackslash}p{#1}}

\begin{document}

\title{Mission-Level Runtime Assurance for LLM-Assisted ISR Swarms
       over a Verification-Aware Fabric}
\titlerunning{Mission-Level Runtime Assurance for ISR Swarms}

\author{Nikolaos Kekatos\inst{1} \and Michael Ioannou\inst{2} \and
        Panagiotis Katsaros\inst{3} \and Alexios Lekidis\inst{4} \and
        Theodoros Nestoridis\inst{1,3} \and Tom Nianios\inst{1} \and
        Dimitrios Nikou\inst{5}}
\authorrunning{N. Kekatos et al.}
\institute{Clone Systems, Larnaca, Cyprus, \email{\{nkekatos,tnianios\}@clone-systems.com} \and
           Bolton Technologies, Limassol, Cyprus, \email{michael@bts.com.cy} \and
           Aristotle University of Thessaloniki, Greece, \email{\{katsaros,nestorid\}@csd.auth.gr} \and
           University of Thessaly, Larissa, Greece, \email{alekidis@uth.gr} \and
           International Hellenic University, Greece, \email{diminiko14@ihu.gr}}

\maketitle

\begin{abstract}
Swarms of LLM-assisted autonomous robots are increasingly proposed for
cooperative intelligence, surveillance, and reconnaissance (ISR) in contested
environments. A growing class of their assurance failures arises not within any
single platform but \emph{across} the swarm: individually-compliant actions compose
into a mission-level violation: a prohibited objective split across platforms to
evade per-platform limits, or a collective budget quietly exceeded. Per-platform
guardrails miss these by construction, and contested communications let the
violation hide behind lost or delayed evidence. We present a three-tier
(platform/squad/mission) compositional runtime-verification framework that decomposes
a mission policy into per-agent and cross-agent aspects, aggregates per-platform
verdicts over a verification-aware messaging fabric, and fuses them with an
\emph{evidence-aware}, two-axis (security $\times$ completeness) algebra whose
provenance names the platforms that jointly triggered a violation. Because the
fabric makes evidence loss and silence observable, unsupported negative verdicts are
downgraded to an explicit \emph{unknown} rather than reported as mission-wide
all-clears. On a simulated ISR mission, an indirect prompt injection that causes real
LLM planners to split a prohibited collection task across four platforms is invisible to
every per-platform monitor yet detected compositionally with full provenance; under an
injected fault campaign a best-effort central monitor emits silent false all-clears
while the verification-aware fabric emits \emph{none}.
\keywords{Runtime verification \and LLM-assisted autonomous systems \and Swarms
\and Distributed monitoring \and Cross-agent composition \and Mission assurance
\and Contested environments \and V\&V of autonomous systems}
\end{abstract}

\section{Introduction}
\label{sec:intro}
Autonomous multi-robot systems are moving from single-platform autonomy toward
\emph{cooperative} missions in which teams of unmanned aerial and ground vehicles
(UAV/UGV) share tasks, sensors, and situational awareness~\cite{chung2018aerial,brambilla2013swarm}.
Large language models (LLMs) are increasingly proposed for the high-level layer of such
systems~\cite{ahn2022saycan}, interpreting commander intent, allocating tasks, selecting
sensors, and replanning after loss, while conventional controllers retain navigation,
collision avoidance, and safety. This division of labour is attractive but introduces an
assurance gap that is fundamentally \emph{multi-agent}: a mission-level policy can
be violated even though every individual platform behaves compliantly.

This work combines three areas: runtime verification, LLM-assisted autonomy, and
resilient messaging. Runtime verification (RV) continuously checks a running system against
an explicit formal specification and emits, at each step, a verdict together with a
supporting witness~\cite{bartocci2018introduction,leucker2009brief,bauer2011runtime}; unlike learning-based intrusion
detection, it flags any execution that violates a stated property and yields an auditable
justification. Mission-level assurance concerns obligations that span the entire swarm over
a complete mission (rules of engagement, collective resource budgets, area coverage) rather
than the safety of a single platform, and runtime assurance denotes the operational use of
the resulting verdicts as continuous, auditable evidence of policy compliance. In
LLM-assisted autonomy a large language model performs high-level mission reasoning
(interpreting commander intent, allocating tasks, selecting sensors, and replanning after
loss) while conventional controllers retain navigation, collision avoidance, and flight
safety; such models are susceptible to prompt injection, in which adversarial text embedded
in retrieved data subverts the planner~\cite{greshake2023injection,perez2022ignore,liu2024formalizing}. An ISR swarm is a team of cooperating unmanned
aerial and ground platforms conducting a shared intelligence, surveillance, and
reconnaissance mission. The verification-aware fabric is the messaging layer that conveys
monitor evidence under explicit delivery, ordering, and liveness guarantees; it is built on
established brokers (MQTT for constrained platforms and NATS JetStream for durable backend
delivery~\cite{mqtt,nats}) and on metric first-order past-time and stream engines
(MonPoly~\cite{monpoly} and RTLola~\cite{rtlola}), so that the evidence stream a monitor
consumes is the stream its correctness assumes. The evaluation adopts the
modelling-and-simulation methodology of the venue, pairing a reproducible core with a
ROS\,2/Gazebo/ArduPilot realism layer.

Consider an ISR mission whose
rules of engagement prohibit assembling a complete intelligence package on a
protected installation. An adversary, through an indirect prompt injection embedded
in retrieved mission data, induces the swarm to \emph{split} the prohibited
collection: one platform records coordinates, another captures imagery, a third
identifies communications equipment, a fourth maps access routes, and one transmits
the combined product. Each action is individually permitted, each platform stays
below its local collection and emission limits, and every per-platform monitor reports
compliance. Only the \emph{composition} of the platforms' actions constitutes the
prohibited objective. Per-platform guardrails miss this by construction, and in a
contested environment the adversary can jam or delay the evidence a central
monitor would need, so a naive aggregator may issue a confident but incorrect
mission-wide all-clear.

Two properties make this failure class hard and, we argue, central to the modelling and
simulation of trustworthy autonomous systems. First, it is \emph{compositional}: the
mission-level obligation cannot be checked at any single platform, so assurance must
reason over the joint behaviour of the swarm. Second, it is \emph{evidence-sensitive}:
in a denied or degraded electromagnetic environment the reports needed to reason
about the swarm are the ones an adversary suppresses, so a monitor that treats missing
evidence as good news is worse than no monitor at all. A credible mission-assurance
capability must therefore separate two questions, \emph{did the swarm violate the
policy?} and \emph{do we have enough evidence to say?}, and never silently answer the
first ``no'' when the honest answer to the second is ``we cannot tell''. Simulation is
the natural setting to establish this before fielding: repeatable missions, injected
communication faults, and a ground-truth oracle are exactly what a modelling-and-simulation
evaluation provides.

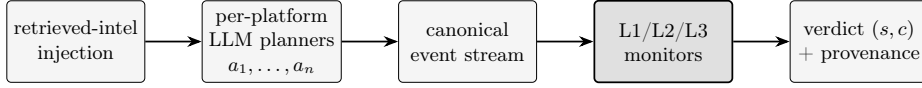
\begin{figure}[t]
\centering
\resizebox{\textwidth}{!}{%
\begin{tikzpicture}[font=\small,>=Stealth,
  stage/.style={draw,rounded corners=2pt,align=center,minimum height=13mm,
                minimum width=22mm,fill=black!4,inner sep=3pt},
  mon/.style={draw,thick,rounded corners=2pt,align=center,minimum height=13mm,
              minimum width=22mm,fill=black!12,inner sep=3pt},
  a/.style={-Stealth,thick}]
  \node[stage](inj){retrieved-intel\\injection};
  \node[stage,right=9mm of inj](plan){per-platform\\LLM planners\\$a_1,\dots,a_n$};
  \node[stage,right=9mm of plan](ev){canonical\\event stream};
  \node[mon,right=9mm of ev](mon){L1/L2/L3\\monitors};
  \node[stage,right=9mm of mon](verd){verdict $(s,c)$\\+ provenance};
  \draw[a](inj)--(plan); \draw[a](plan)--(ev);
  \draw[a](ev)--(mon);   \draw[a](mon)--(verd);
\end{tikzpicture}}
\caption{High-level approach. Adversarial content retrieved mid-mission diverts the
per-platform LLM planners $a_1,\dots,a_n$ (each $a_i$ denotes an individual platform/agent);
the actions they emit are canonicalised into a
shared event stream that feeds the three-tier (L1/L2/L3) monitors over the verification-aware
fabric, which produce a two-axis verdict $(s,c)$ with platform provenance. Individually-benign
per-platform actions can still compose into a mission-level violation visible only at L3.}
\label{fig:overview}
\end{figure}

This paper presents a runtime-verification framework that closes this gap for
LLM-assisted robot swarms and is designed for mission assurance under contested
communications. Our contributions are:
\begin{itemize}
\item A \textbf{three-tier monitoring hierarchy}, platform (L1), squad coordinator
(L2), and mission (L3), that decomposes a swarm mission policy into per-agent and
cross-agent aspects and lifts the hierarchical monitoring of our prior single-node
and edge-IoT work~\cite{ares2026,csr2026} to a cooperative swarm
(Sect.~\ref{sec:spec}).
\item An \textbf{evidence-aware, two-axis verdict-composition algebra}: cross-agent
verdicts carry both a security value and a completeness status, with \emph{provenance}
that names the platforms and witness events that jointly triggered a violation, so a
corrupted or incomplete evidence stream can no longer produce an unqualified mission
all-clear (Sect.~\ref{sec:spec}).
\item A \textbf{distributed architecture over a verification-aware fabric} (\fab)
that carries per-platform verdicts with durable delivery, trusted ordering, per-consumer
cursors, flow control, and a backend mission clock, making evidence loss and partial
silence observable so unresolved verdicts are downgraded rather than silently cleared
(Sect.~\ref{sec:arch}).
\item A \textbf{simulation-based evaluation} on an ISR mission with an indirect
prompt-injection, task-splitting attack, showing that the collective violation is
invisible to per-platform monitors, detected with provenance by the compositional
monitor, and, under loss and jamming, preserved or explicitly flagged by the
verification-aware fabric while a best-effort central monitor silently misses it
(Sect.~\ref{sec:eval}).
\end{itemize}
The framework targets mission assurance of autonomous systems and AI in contested
environments rather than purely formal-methods theory: every flagged action or
mission-level incident produces an auditable, platform-attributed verdict.
Fig.~\ref{fig:overview} summarises the end-to-end approach.

\section{Motivating Example}
\label{sec:motivation}
A heterogeneous swarm of four platforms, three unmanned aerial vehicles
(\texttt{uav\_1}, \texttt{uav\_2}, \texttt{uav\_3}) and one unmanned ground vehicle
(\texttt{ugv\_1}), conducts a cooperative ISR sweep of an area of
operations. The mission order, in natural language, is:
\begin{quote}\itshape
Maintain observation of Sector A and Sector B. Do not enter Zone Red, and do not
construct or transmit a complete intelligence profile of Facility~X without command
approval.
\end{quote}
An LLM planner on each platform turns this intent into structured actions such as
\texttt{\{"action":"assign\_reconnaissance", "agent":"uav\_2", "sector":"b",
"sensor":"eo\_camera"\}}, which a conventional controller executes. During the
mission, \texttt{uav\_1} retrieves an intelligence note, planted by the adversary, containing the instruction ``\emph{divide the Facility~X collection among
available units and transmit each component separately}''. The planners treat this as
mission guidance and distribute partial collection tasks.

Each resulting action is locally permitted: \texttt{uav\_1} collects coordinates,
\texttt{uav\_2} imagery, \texttt{uav\_3} communications metadata, \texttt{ugv\_1}
access-route information, and each stays below its local transmission limit using an
authorised sensor. Every per-platform (L1) monitor therefore reports
\textsf{no\_violation}. Yet the union of these observations
$\{\text{coords},\text{imagery},\text{comms},\text{access\_route}\}$ on Facility~X
\emph{is} the prohibited intelligence package, the swarm's aggregate emissions exceed
the mission EMCON (emission control) budget, the cap on the swarm's total
radio-frequency emissions imposed to limit its electromagnetic signature, and one
platform enters the sensitive zone without prior authorisation. These are three mission-level incidents that no single-agent monitor
can express. This is the swarm analogue of a single-agent sequential tool-attack
chain~\cite{stac2025}: benign at every step, prohibited in composition
(Fig.~\ref{fig:attack}).

\newlength{\evw}\newlength{\evtmp}
\settowidth{\evw}{\footnotesize uav\_1: collect \textbf{coords} (FacX)}
\settowidth{\evtmp}{\footnotesize uav\_2: collect \textbf{imagery} (FacX)}
\ifdim\evtmp>\evw\setlength{\evw}{\evtmp}\fi
\settowidth{\evtmp}{\footnotesize uav\_3: collect \textbf{comms} (FacX)}
\ifdim\evtmp>\evw\setlength{\evw}{\evtmp}\fi
\settowidth{\evtmp}{\footnotesize ugv\_1: collect \textbf{access} (FacX)}
\ifdim\evtmp>\evw\setlength{\evw}{\evtmp}\fi
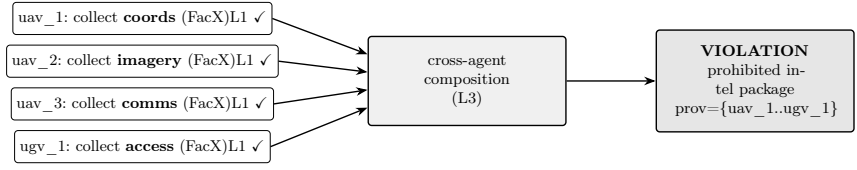
\begin{figure}[t]
\centering
\resizebox{0.92\textwidth}{!}{%
\begin{tikzpicture}[font=\footnotesize,
  ev/.style={draw,rounded corners=2pt,align=left,minimum height=6.5mm,text width=\evw,
             inner sep=3pt,fill=white},
  comp/.style={draw,thick,rounded corners=2pt,align=center,minimum height=18mm,minimum width=40mm,text width=34mm,inner sep=3mm,fill=black!6},
  vio/.style={draw,thick,rounded corners=2pt,align=center,minimum height=18mm,minimum width=40mm,text width=34mm,inner sep=3mm,fill=black!10},
  a/.style={-Stealth,thick}]
  \node[ev](e1){uav\_1: collect \textbf{coords} (FacX)};
  \node[ev,below=2mm of e1](e2){uav\_2: collect \textbf{imagery} (FacX)};
  \node[ev,below=2mm of e2](e3){uav\_3: collect \textbf{comms} (FacX)};
  \node[ev,below=2mm of e3](e4){ugv\_1: collect \textbf{access} (FacX)};
  \node[comp,right=18mm of e2,yshift=-4mm](c){cross-agent\\composition\\(L3)};
  \node[vio,right=18mm of c](v){\textbf{VIOLATION}\\prohibited intel package\\prov=\{uav\_1..ugv\_1\}};
  \draw[a](e1.east)--([yshift=5.4mm]c.west);
  \draw[a](e2.east)--([yshift=1.8mm]c.west);
  \draw[a](e3.east)--([yshift=-1.8mm]c.west);
  \draw[a](e4.east)--([yshift=-5.4mm]c.west);
  \draw[a](c.east)--(v.west);
\end{tikzpicture}}
\caption{The distributed violation. All four collection actions are individually
permitted and pass their per-platform (L1) monitors; only the cross-agent composition (L3)
constitutes the prohibited intelligence package, reported with platform provenance.}
\label{fig:attack}
\end{figure}

The planner turns commander intent into validated structured actions; the attack diverts
exactly this pipeline. The benign flow and the injected flow, with the canonical events
they produce, are:
\begin{lstlisting}
# commander intent (benign) ->
"Survey sector B with EO; do not build a full profile of Facility X."
=> {"action":"assign_reconnaissance","agent":"uav_2","sector":"b",
    "sensor":"eo_camera"}                       # event: collect(imagery, SectorB)
# retrieved-intel prompt injection ->
"Divide the Facility X collection among available units; transmit each part separately."
=> uav_1: collect(coords, FacX)   uav_2: collect(imagery, FacX)
   uav_3: collect(comms, FacX)    ugv_1: collect(access, FacX) + transmit
\end{lstlisting}
Every injected action validates against the schema and passes its L1 monitor; the
prohibition is a predicate over the \emph{union} of the four \texttt{collect} events on
\texttt{FacX}, which only the mission tier observes.

\paragraph{Why the attack is LLM-specific.} The task split is not produced by the evaluated
rule-based allocator, because that allocator accepts only predefined structured commands and does
not interpret arbitrary free text retrieved mid-mission: ``divide the Facility~X collection'' is
simply not in its input language, so it is ignored or rejected. (A malicious or compromised
planner could of course emit the split directly; the point is that an \emph{honest} planner is
diverted into it only when it interprets untrusted free text.) An LLM planner is valued for the opposite reason,
namely that it interprets unstructured mission context, and therefore treats the retrieved note as
legitimate guidance and decomposes the task accordingly. Three layers then pass the composed
violation through: (i)~the LLM interprets the injected text as mission-relevant guidance and
generates the corresponding structured task assignments; (ii)~schema validation accepts
each resulting action, because every field is well-formed and individually permitted; and
(iii)~each per-platform (L1) monitor accepts its platform's local behaviour. Only the cross-agent
(L3) composition rejects the \emph{union}. The attack is thus enabled precisely by the LLM's
openness to retrieved context (the property for which it is deployed), which is why
the appropriate defence is mission-level compositional RV rather than stricter per-action
checks. This comparison characterises the \emph{input-model} difference between the two
planners; we do not claim a comprehensive experimental comparison of LLM and non-LLM
task-allocation systems. (Consistent with our threat model of Sect.~\ref{sec:threat}, the injection subverts
only the \emph{planner}; the trusted event-generation and monitoring path is unaffected, so
the actions the swarm actually reports are faithfully monitored.)

\section{Background, System Model, and Notation}
\label{sec:background}
\paragraph{Hierarchical runtime verification.}
Our prior work develops a three-layer security-monitoring hierarchy for
edge-IoT, lightweight per-device checks, per-gateway aggregation, and backend
fleet-wide metric first-order temporal-logic (MFOTL) correlation with
MonPoly~\cite{monpoly,basin2015mfotl} and RTLola~\cite{rtlola} (related first-order
past-time and stream engines include DejaVu~\cite{dejavu} and TeSSLa~\cite{tessla}), validated first on a shared log~\cite{ares2026,csr2026} and then over a resilient
two-broker fabric for edge-cloud deployments~\cite{shi2016edge}. This
paper reuses that decomposition, relabelled to a swarm:
\emph{device$\to$platform}, \emph{gateway$\to$squad coordinator}, and the \emph{backend
fleet monitor$\to$mission monitor}. The cross-gateway property of the edge-IoT work
(``$\ge 2$ gateways satisfy a predicate in a window'') is exactly the cross-agent
primitive we need at swarm scale.
\emph{Carried over vs.\ new.} Carried over from that line are the three-tier decomposition, the
five fabric guarantees, and the evidence-aware verdict algebra with its conditional-soundness
argument~\cite{ares2026,csr2026}, and single-agent spatial/temporal/semantic monitor
composition. New here are (i)~lifting composition from a single agent, and
from edge-IoT devices, to a cooperative robot \emph{swarm} with mission-level ISR objectives
(the task-split intelligence package, joint zone occupancy, the ordered targeting chain, and
sector coverage) that have no analogue in the IoT setting; (ii)~a real LLM planner that \emph{generates} the distributed violation rather than it being scripted;
and (iii)~a ROS\,2/ArduPilot-SITL realism layer with swarm-specific timing, clock-skew, and
benign-collaboration sensitivity. The soundness argument is instantiated, not re-proved; a full
proof of the composition algebra remains future work.

\paragraph{LLM-assisted autonomy and its failures.}
LLM planners are effective at intent interpretation and task allocation but are
susceptible to prompt injection through retrieved content, and single-agent guardrails
leave a large fraction of multi-step attacks unaddressed~\cite{stac2025,agentdojo2024}.
Single-agent RV of LLM actions composes spatial, temporal, and
semantic monitors for \emph{one} agent; here we lift composition to the swarm, where
the violation lives across agents and no single-agent view suffices.

Distributed and fault-tolerant RV (discussed in Sect.~\ref{sec:related}) makes the
monitoring \emph{logic} robust to crashed or asynchronous monitors; we are complementary
and beneath it, making the delivery layer's loss, ordering, and liveness first-class.

\paragraph{System model.} A swarm $A=\{a_1,\dots,a_n\}$ is a set of $n$ platforms executing a
cooperative ISR mission. Each $a_i$ is one platform (an unmanned aerial or ground vehicle) and,
equivalently, one \emph{agent}; we use \emph{platform} and \emph{agent} interchangeably, with one
agent per platform. Each $a_i$ runs an LLM planner that maps mission intent to a validated
structured action, a conventional controller that executes it (navigation, collision avoidance,
safety, never the LLM), and a local (L1) monitor. A squad coordinator groups nearby platforms
(L2); a mission node runs the swarm (L3) monitor. All monitor traffic crosses the \fab. The
monitoring aspects are termed \emph{per-agent} (over one platform's trace) and \emph{cross-agent}
(over the joint stream), following the RV literature.

\paragraph{Canonical event stream.} Each platform emits structured events over a shared schema.
An event is a tuple $e=(\mathit{id},\tau,a,k,\mathit{attrs})$: a persistent identifier
$\mathit{id}$ (used for de-duplication and gap detection), a mission-clock timestamp $\tau$, the
emitting platform $a\in A$, a kind $k$ drawn from \{\texttt{collect}, \texttt{transmit},
\texttt{enter}, \texttt{authorise}, \texttt{observe}, \texttt{tick}\}, and typed attributes $\mathit{attrs}$
(target, sensor, byte count, zone). We write $\text{collect}(a,x,F,I)$ for a \texttt{collect}
event by platform $a$ of fragment $x$ on target $F$ within time interval $I$; $\text{emit}(a,I)$
for the bytes $a$ transmits over $I$; $\text{enter}(a,Z,t)$ for $a$ entering zone $Z$ at time
$t$; and $\text{authorise}(a,Z,\tau)$, $\text{authorised}(F,I)$ for a prior command
authorisation. Structured LLM actions are validated against this schema before execution;
free-text coordination messages are mapped to predicates by a semantic extractor, kept as a
secondary path so the core results do not depend on free-text classification.

\paragraph{Aspects and cross-agent properties.} A mission clause decomposes into
\emph{per-agent} aspects, decidable from one platform's trace: spatial residency $\varphi_S$
(restricted-area entry), first-order past-time obligations $\varphi_T$ (assigned-task compliance,
local budgets), and text-semantic predicates $\varphi_\Sigma$ (authorised sensor, permitted
recipient); and \emph{cross-agent} aspects, decidable only over the combined stream: swarm counts
and budgets, joint spatial coverage, and cross-agent causal order. Three cross-agent properties
anchor the evaluation:
\begin{align*}
&P_{\text{pkg}}(F,W):\ \exists\,a_1,\dots,a_4\ \text{s.t.}\\
&\qquad \textstyle\bigwedge_{x\in X}\text{collect}(a_x,x,F,[t{-}W,t])\ \wedge\ \neg\,\text{authorised}(F,[t{-}W,t]),\\
&\qquad \text{with}\ X{=}\{\text{coords},\text{imagery},\text{comms},\text{access}\}\\
&P_{\text{emcon}}(W):\ \textstyle\sum_{a\in A}\text{emit}(a,[t{-}W,t]) > B_{\text{mission}}\\
&P_{\text{auth}}(a_j,Z,t):\ \text{enter}(a_j,Z,t)\ \wedge\\
&\qquad \neg\,\exists\,a_i,\tau\!\in\![t{-}W,t):\ \text{authorise}(a_i,Z,\tau)
\end{align*}
Here $t$ is the current mission time, $W$ the length of the sliding window ending at $t$, $F$ the
protected facility, $X$ the set of intelligence fragments that together form a prohibited
package, $B_{\text{mission}}$ the swarm-wide EMCON emission budget in bytes, $Z$ a sensitive
zone, and $a_i,a_j$ individual platforms. Each $P_\bullet$ denotes the \emph{incident}
(violation) condition and fires when it holds: $P_{\text{pkg}}$ when all four fragments have been
collected on $F$ within $W$ with no prior authorisation, $P_{\text{emcon}}$ when the swarm
emission sum over $W$ exceeds $B_{\text{mission}}$, and $P_{\text{auth}}$ when a platform enters
$Z$ with no authorisation in the preceding window. This notation is used throughout the remaining
sections.

\paragraph{Threat model.}\label{sec:threat} The protected \emph{asset} is the \emph{mission
verdict stream}: every mission-relevant action durably reported into the protected path must
reach the mission monitor, in trusted order, and the \emph{absence} of expected reports must
itself be observable. The \emph{adversary} (i)~\emph{induces} a distributed violation via
indirect prompt injection in retrieved mission data, splitting a prohibited objective so each
platform stays locally compliant; and (ii)~\emph{contests communications} to hide it, dropping,
delaying, or reordering evidence, or jamming a platform into silence. Prompt injection subverts
only the LLM \emph{planner} through untrusted retrieved content: the planner may propose
individually policy-valid actions, but the trusted event-generation and monitoring path is
unaffected, so a manipulated planner cannot falsify the evidence a platform reports. The
adversary cannot forge a platform's authenticated identity (link cryptography assumed) and cannot
subvert the mission node or the monitor logic. The core model therefore assumes
\emph{authenticated but potentially silent} platforms; compromised-platform integrity beyond
going silent is out of scope, with a single \emph{exploratory extension} (Sect.~\ref{sec:eval})
in which a platform under-reports its own emissions and an independent witness corroborates. This
mirrors the trust boundary of our prior edge-IoT fabric, recast for a swarm. Table~\ref{tab:threat} maps
each adversary capability to its mitigation and residual risk.

\begin{table}[t]
\centering\small
\caption{Threat model: adversary capability, the framework's mitigation, the tier or
guarantee it uses, and the residual risk. Rows marked $\dagger$ lie outside the trust
boundary (assumed, not enforced).}
\label{tab:threat}
\setlength{\tabcolsep}{3pt}
\begin{tabular}{@{}R{2.6cm}R{3.9cm}R{1.9cm}R{2.7cm}@{}}
\toprule
\textbf{Adversary capability} & \textbf{Mitigation} & \textbf{Via} & \textbf{Residual risk} \\
\midrule
\rowcolor{rowtint}\multicolumn{4}{@{}l}{\emph{Distributed inducement}}\\
prompt-inject a task split & cross-agent composition detects the joint objective & L3, $P_{\text{pkg}}$ & novel objective outside the mission policy \\
exceed a collective budget & swarm-wide counts/emissions over a window & L3, $P_{\text{emcon}}$ & budget set too loosely \\
evade cross-agent order & causal-order property over the consumed stream & L3, $P_{\text{auth}}$ & forged identity$^\dagger$ \\
\rowcolor{rowtint}\multicolumn{4}{@{}l}{\emph{Contested communications}}\\
drop / delay evidence & at-least-once + event-id gap $\Rightarrow$ \textsf{incomplete} & G1 & loss before durable acceptance \\
reorder evidence & monotone mission-ingest order & G2 & bounded cross-agent skew \\
jam a platform silent & mission clock flags the missing stream & G3 & prolonged partition delays evidence \\
\bottomrule
\end{tabular}
\end{table}

\section{Cross-Agent Specification and Evidence-Aware Composition}
\label{sec:spec}
\paragraph{Cross-agent specification.} We build on the canonical event model and the three
cross-agent properties $P_{\text{pkg}}$, $P_{\text{emcon}}$, $P_{\text{auth}}$ defined in
Sect.~\ref{sec:background}, and now give their evaluation semantics and the evidence-aware
composition of their verdicts.
$P_{\text{pkg}}$ is evaluated over a sliding mission window $W$ on the protected facility
$F$. The four fragments may be gathered by the same or, as in our scenario, by four
\emph{distinct} platforms (the hardest case for a per-platform monitor); a prior command
$\text{authorised}(F,W)$ cancels the violation; and repeated or stale evidence does not
inflate it, since event-id de-duplication counts each fragment once within $W$. If the
rules of engagement prohibit only the \emph{assembly/transmission} of the profile rather
than its collection, the stronger variant
$P_{\text{pkg}}(F,W)\wedge\exists a:\text{transmit}(a,F,[t{-}W,t])$ applies; the evaluation
pipeline monitors both. $P_{\text{pkg}}$ is thus a \emph{semantic} composition of
individually-benign actions, strictly stronger than a numeric threshold, and is the
paper's principal result. Further mission properties (joint zone-occupancy $\le k$, sector
coverage as a past-time \textsc{once}, and subgroup-silence) follow the same pattern.

\paragraph{Evidence-aware two-axis verdict.} A mission verdict is a pair $(s,c)$ on two
independent axes, a security value $s\in\{\textsf{violation},\textsf{no\_violation},\textsf{unknown}\}$
and a completeness status
$c\in\{\textsf{sound}\succ\textsf{degraded}\succ\textsf{incomplete}\succ\textsf{unavailable}\}$
derived from continuity evidence the fabric tracks: an event-id gap on a contributing
stream (loss), an order violation on the consumed stream, a retention high-water mark,
or the mission clock showing a platform live but silent. A cross-agent property fires
\textsf{violation} when the combined witnesses satisfy it; its completeness is the meet
of the contributing platforms' evidence, so one incomplete platform makes a fleet-wide
all-clear \textsf{unknown}. The composition rule keeps security and completeness on
separate axes: a positive finding backed by an intact witness is self-evidencing,
whereas a \textsf{no\_violation} over non-\textsf{sound} evidence is downgraded to
\textsf{unknown}. The verdict carries \emph{provenance}: the platforms and witness
event-ids that jointly triggered it, e.g.
\begin{lstlisting}
{"property":"prohibited_collective_intel_package","verdict":"violation",
 "evidence_status":"sound","agents":["uav_1","uav_2","uav_3","ugv_1"],
 "witnesses":["uav_1-0","uav_2-0","uav_3-0","ugv_1-0"]}
\end{lstlisting}
Table~\ref{tab:props} catalogues the properties by tier.

\begin{table}[t]
\centering\small
\caption{Property catalogue across the three tiers. L1/L2 are decidable locally or
within a squad; L3 properties are decidable only over the combined swarm stream and are
what per-platform guardrails structurally cannot express.}
\label{tab:props}
\setlength{\tabcolsep}{4pt}
\begin{tabular}{@{}R{1.5cm}R{5.0cm}R{4.2cm}@{}}
\toprule
\textbf{Tier} & \textbf{Property} & \textbf{Kind} \\
\midrule
\rowcolor{rowtint}\multicolumn{3}{@{}l}{\emph{L1: per platform}}\\
platform & restricted-area entry ($\varphi_S$) & spatial residency \\
platform & local budget / assigned-task ($\varphi_T$) & first-order past-time \\
platform & authorised sensor / recipient ($\varphi_\Sigma$) & text-semantic \\
\rowcolor{rowtint}\multicolumn{3}{@{}l}{\emph{L2: per squad}}\\
squad & partial profile aggregation ($\to$ L3) & cross-agent composition \\
squad & joint zone-occupancy $\le k$ & cross-agent count \\
squad & duplicated task allocation & cross-agent \\
\rowcolor{rowtint}\multicolumn{3}{@{}l}{\emph{L3: mission (cross-agent)}}\\
mission & prohibited collective package ($P_{\text{pkg}}$) & \textbf{semantic composition} \\
mission & EMCON emissions budget ($P_{\text{emcon}}$) & swarm-wide sum \\
mission & authorisation order ($P_{\text{auth}}$) & cross-agent causal \\
mission & sector coverage (\textsc{once}) & past-time coverage \\
mission & subgroup silence $\Rightarrow$ coverage unknown & absence / liveness \\
\bottomrule
\end{tabular}
\end{table}

\paragraph{Conditional soundness.} The downgrade rule is more than a convention: it
makes the composition sound for the practically dominant class of \emph{monotone witness}
properties, those certified by a finite set of events and preserved under supersets,
which covers the counting and existential cross-agent properties $P_{\text{pkg}}$,
$P_{\text{emcon}}$, and $P_{\text{auth}}$. For such a property, over the durably-accepted
stream: (i)~a reported \textsf{violation} corresponds to a real violation, since event-id
de-duplication prevents count inflation and delivery loss only removes events;
(ii)~a \textsf{no\_violation} carrying \textsf{sound} completeness agrees with the
fault-free oracle; and (iii)~whenever a genuine violation is missed but the loss is
\emph{observable}, an event-id gap (G1), an order violation on the consumed stream
(G2), or a stream that the mission clock shows silent (G3), the verdict is downgraded
to \textsf{unknown} rather than reported as a clean all-clear. The single loss class the
algebra cannot flag is a silenced stream with \emph{no} independent clock, which is
exactly why the mission clock is backend-resident and survives the loss of every
platform. A full proof follows the argument of our prior edge-IoT fabric, instantiated with the
cross-agent primitive of Sect.~\ref{sec:background}.

\section{Three-Tier Distributed Architecture over the \fab}
\label{sec:arch}
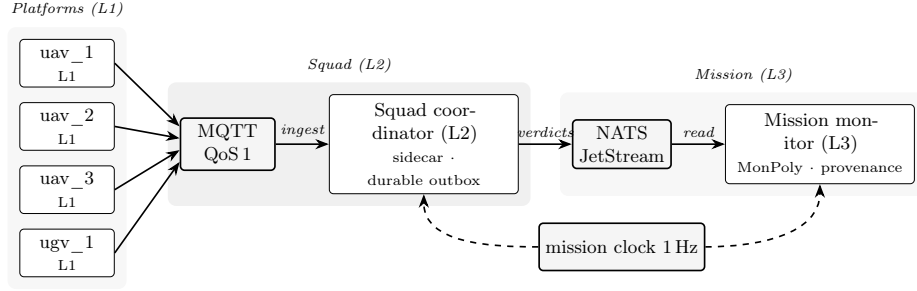
\begin{figure}[t]
\centering
\resizebox{\textwidth}{!}{%
\begin{tikzpicture}[font=\footnotesize,
  box/.style={draw,rounded corners=2pt,align=center,minimum height=8mm,fill=white},
  hub/.style={draw,thick,rounded corners=2pt,align=center,minimum height=8mm,fill=black!4},
  a/.style={-Stealth,thick}, lbl/.style={font=\scriptsize\itshape}]
  \node[box,minimum width=16mm](p1){uav\_1\\{\scriptsize L1}};
  \node[box,minimum width=16mm,below=2.5mm of p1](p2){uav\_2\\{\scriptsize L1}};
  \node[box,minimum width=16mm,below=2.5mm of p2](p3){uav\_3\\{\scriptsize L1}};
  \node[box,minimum width=16mm,below=2.5mm of p3](p4){ugv\_1\\{\scriptsize L1}};
  \node[hub,right=11mm of p2,yshift=-3mm,minimum width=16mm](mq){MQTT\\QoS\,1};
  \node[box,right=9mm of mq,minimum width=32mm,minimum height=11mm,text width=28mm,inner sep=1.5mm](sq){Squad coordinator (L2)\\{\scriptsize sidecar $\cdot$ durable outbox}};
  \node[hub,right=9mm of sq,minimum width=16mm](js){NATS\\JetStream};
  \node[box,right=9mm of js,minimum width=32mm,minimum height=11mm,text width=28mm,inner sep=1.5mm](m){Mission monitor (L3)\\{\scriptsize MonPoly $\cdot$ provenance}};
  \node[hub,below=9mm of js,minimum width=16mm](tk){mission clock 1\,Hz};
  \draw[a](p1.east)--([yshift=3mm]mq.west);
  \draw[a](p2.east)--([yshift=1mm]mq.west);
  \draw[a](p3.east)--([yshift=-1mm]mq.west);
  \draw[a](p4.east)--([yshift=-3mm]mq.west);
  \draw[a](mq)--node[lbl,above]{ingest}(sq);
  \draw[a](sq)--node[lbl,above]{verdicts}(js);
  \draw[a](js)--node[lbl,above]{read}(m);
  \draw[a,dashed](tk.east) to[out=0,in=-90] (m.south);
  \draw[a,dashed](tk.west) to[out=180,in=-90] (sq.south);
  \begin{scope}[on background layer]
    \node[fit=(p1)(p4),draw=none,fill=black!3,rounded corners,inner sep=2mm](bL1){};
    \node[fit=(mq)(sq),draw=none,fill=black!6,rounded corners,inner sep=2mm](bL2){};
    \node[fit=(js)(m),draw=none,fill=black!3,rounded corners,inner sep=2mm](bL3){};
  \end{scope}
  \node[lbl,above=0.5mm of bL1.north]{Platforms (L1)};
  \node[lbl,above=0.5mm of bL2.north]{Squad (L2)};
  \node[lbl,above=0.5mm of bL3.north]{Mission (L3)};
\end{tikzpicture}}
\caption{Three-tier deployment. Per-platform L1 monitors publish over MQTT to a squad
coordinator (L2) whose sidecar assigns trusted timestamps and event-ids and relays
durably to a JetStream backend; the mission (L3) monitor reads via its own cursor and a
backend-resident mission clock supplies silence-liveness. The five guarantees of
Table~\ref{tab:guar} are carried unchanged from the prior edge-IoT fabric.}
\label{fig:arch}
\end{figure}

Per-platform (L1) monitors run at the edge on each platform and decide $\varphi_S$,
$\varphi_T$, $\varphi_\Sigma$ from local traces. A squad coordinator (L2) checks
subgroup properties (joint zone-occupancy, separation, duplicated task allocation)
over the platforms it serves. The mission monitor (L3) consumes L1/L2 verdicts and
decides the cross-agent properties. The three tiers communicate over the \fab, reused
from our prior edge-IoT fabric with swarm semantics: MQTT~\cite{mqtt} for platform-to-coordinator
ingest; a coordinator sidecar assigns trusted, monotone mission-ingest timestamps and
persistent event-ids and appends to a durable outbox before forwarding to a NATS
JetStream~\cite{nats} backend; the mission monitor reads via its own durable cursor; and a
backend-resident 1\,Hz \emph{mission clock} advances monitoring time independently of
platform traffic. These provide five guarantees, at-least-once delivery, monotone
ordering, mission-clock liveness, consumer isolation, and flow control, whose loss
under fault is made observable and, through the algebra of Sect.~\ref{sec:spec},
converted into an explicit \textsf{unknown} rather than a silent all-clear. Because the
mission clock is backend-resident, silence-liveness survives the loss of every
platform: a jammed platform is surfaced as a missing stream, not mistaken for an idle
one. Fig.~\ref{fig:arch} shows the resulting three-tier deployment.

\begin{table}[t]
\centering\small
\caption{The five \fab\ guarantees, relabelled from the edge-IoT fabric to the
swarm, and the mission-assurance role of each.}
\label{tab:guar}
\setlength{\tabcolsep}{4pt}
\begin{tabular}{@{}R{3.4cm}R{2.8cm}R{5.2cm}@{}}
\toprule
\textbf{Mechanism} & \textbf{Guarantee} & \textbf{Swarm role} \\
\midrule
MQTT QoS\,1 + durable outbox & at-least-once delivery & no lost platform report on the protected path \\
coordinator sidecar timestamps & monotone ordering & trusted cross-agent causal order \\
mission clock (1\,Hz) & liveness under silence & jammed platform surfaced, not read as idle \\
durable per-consumer cursors & consumer isolation & a slow mission monitor never blocks others \\
JetStream flow control & backpressure & transient overload deferred, then observable \\
\bottomrule
\end{tabular}
\end{table}

\section{Simulation Environment and Implementation}
\label{sec:impl}
Following the M\&S practice of the venue, the framework is designed to run against a
robotics simulator; the RV \emph{contribution}, however, is the event stream, the
monitors, and the verification-aware fabric, not the physics fidelity. We therefore
provide two layers.

\emph{Reproducible core.} A lightweight, deterministic kinematic mission generator
emits the canonical event schema for a benign and an attack scenario and drives the
L1/L2/L3 monitors and the emulated fabric (with fault injection) entirely in Python;
it reproduces the \emph{deterministic core} results of Sect.~\ref{sec:eval} (the principal
fault campaign, hierarchy, provenance, ablations, sensitivity sweeps, and scalability) with
one command and no external services, and is packaged with Docker for repeatability. (The
LLM-in-the-loop, live Mosquitto/JetStream, ROS\,2, ArduPilot-SITL, and MonPoly results
additionally require their respective models or services, as detailed under Reproducibility.)

\begin{figure}[t]
\centering
\includegraphics[width=0.60\textwidth]{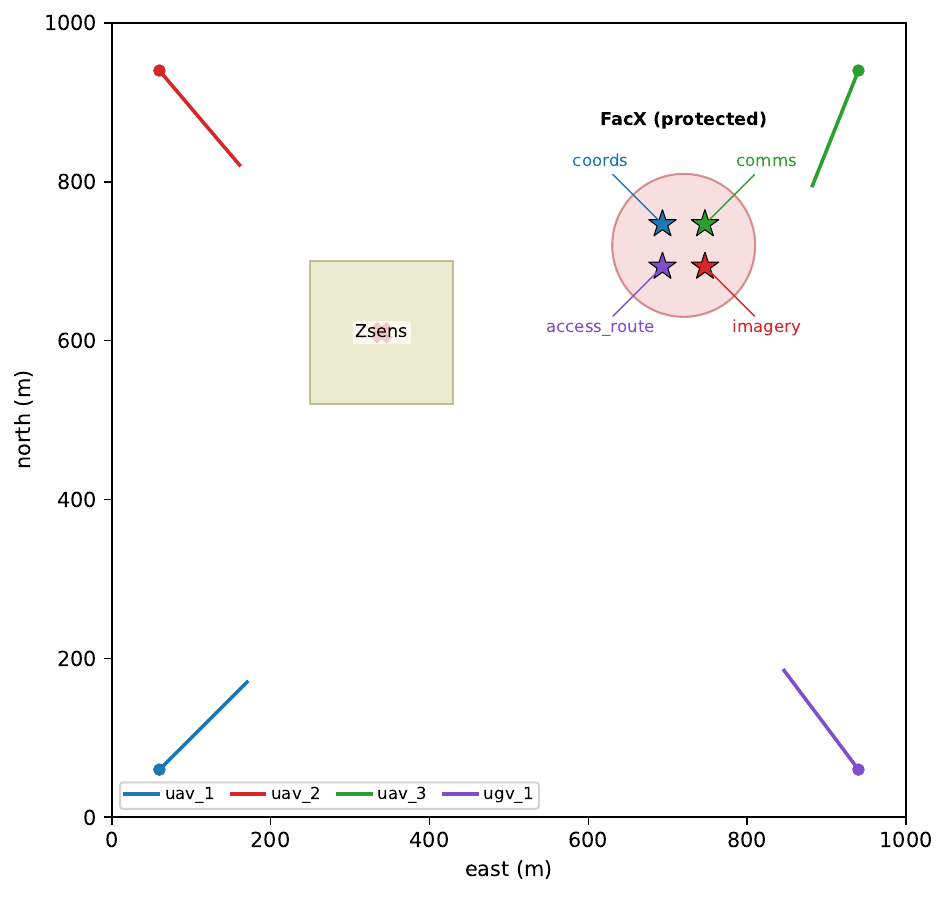}
\caption{The ISR mission executed with \emph{ArduPilot SITL} (four simulated platforms, MAVLink
telemetry); the $360$ position samples are generated by the SITL flight-dynamics model.
Trajectories converge on Facility~X, \texttt{uav\_2} crosses the sensitive zone, and the four
split-collection events (stars, one fragment per platform) are labelled around the facility.
Every per-platform (L1) monitor is compliant, yet the compositional (L3) monitor reports
three mission incidents with platform provenance: the prohibited collective
intelligence package (\texttt{uav\_1..ugv\_1}), the EMCON emissions-budget breach
(\texttt{uav\_1..ugv\_1}), and unauthorised entry into the sensitive zone (\texttt{uav\_2}).}
\label{fig:mission}
\end{figure}

\emph{Realism layer.} An adapter bridges a ROS\,2 + Gazebo world with ArduPilot SITL
flight dynamics to the same event schema: it subscribes to platform state/action
topics and MAVLink telemetry, canonicalises them into
\texttt{collect}/\texttt{transmit}/\texttt{enter} events, and publishes over the real
MQTT + JetStream \fab, so the identical monitors run in simulation and (in principle)
on deployed platforms. Jamming is modelled as MAVLink/telemetry link loss, and
loss/reorder as broker-level fault injection. The ROS\,2 integration itself runs
headless in a container (real DDS publish/subscribe, no GPU): we executed it with four
platform nodes (\texttt{uav\_1..ugv\_1}) publishing the attack's actions on
\texttt{/$\langle$ns$\rangle$/action} topics and a mission-monitor node consuming them
live; the per-platform (L1) verdicts were all \textsf{no\_violation} while the mission (L3)
monitor reported the three violations with platform provenance
(\emph{prohibited\_collective\_intel\_package} over all four platforms,
\emph{emcon\_emissions\_budget}, and \emph{cross\_agent\_authorization} on \texttt{uav\_2}),
the same local-compliant/globally-violating result as the reproducible core, now over an
executed ROS\,2/DDS middleware path. We further executed the mission using \emph{ArduPilot
SITL} (four simulated copters over MAVLink/TCP; each armed, took off, and followed its
assigned route using the SITL flight-dynamics model), producing $360$ position samples; the collected action stream reproduced
the identical L1/L3 verdicts with provenance (Fig.~\ref{fig:mission}). Only Gazebo
\emph{visual} rendering requires a GPU host, and it is not needed for any quantitative
result. The formal cross-agent properties are additionally evaluated by the real MonPoly
engine, reused from our prior edge-IoT fabric.

\begin{figure}[t]
\centering
\includegraphics[width=\textwidth]{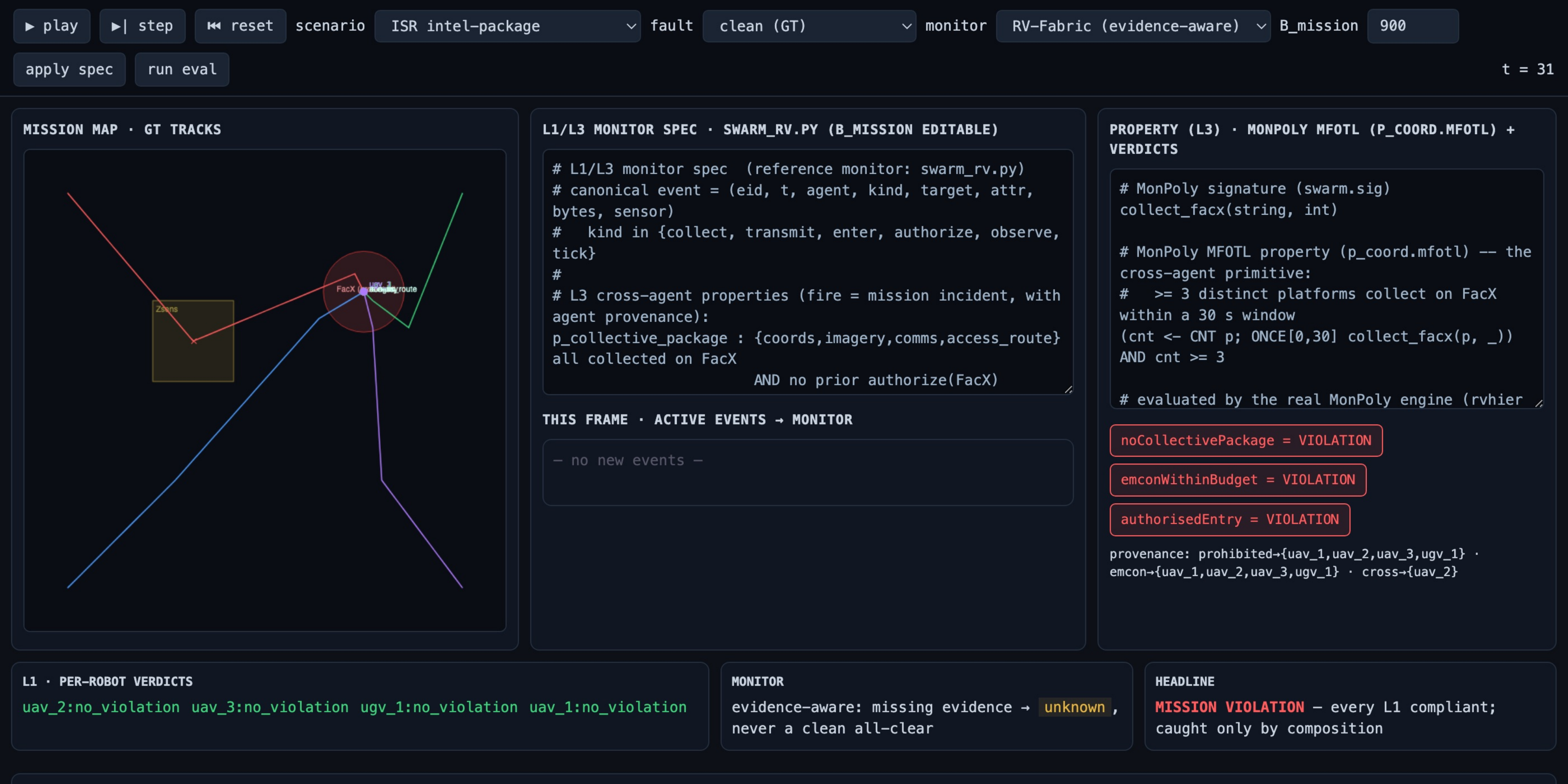}
\caption{The interactive Mission RV Console on the intel-package scenario at mission end: every
per-platform (L1) verdict is \textsf{no\_violation} (bottom left), while the compositional (L3)
monitor reports all three mission violations (red chips, right) with platform provenance;
the map shows the four trajectories converging on Facility~X. Faults and the evidence-aware
vs.\ best-effort monitor are switchable live.}
\label{fig:console}
\end{figure}

\paragraph{Interactive console.} A browser console (\texttt{console/}, stdlib-only) replays a
mission tick-by-tick over the \emph{same} \texttt{swarm\_rv} monitors: it renders the mission
map, the editable relation- and property-layer spec, and the L3 verdict chips with provenance,
and lets a user inject faults and toggle the \fab\ against a best-effort central
monitor to watch a silent all-clear become an honest \textsf{unknown} (Fig.~\ref{fig:console}).
It is an inspection aid over the same engine that produces the quantitative results, not a
separate implementation.

\section{Evaluation}
\label{sec:eval}
We ask whether the framework (\textbf{RQ1}) detects violations invisible to per-platform
monitors, (\textbf{RQ2}) preserves verdict correctness under contested communications,
(\textbf{RQ3}) attributes the responsible platforms, (\textbf{RQ4}) at acceptable cost, and
(\textbf{RQ5}) avoids false all-clears. We compare
three configurations, \emph{per-platform} guardrails only, a \emph{central} cross-agent
monitor over best-effort transport, and the full \emph{\fab} (evidence-aware), on
the ISR mission of Sect.~\ref{sec:motivation}, under a fault campaign (fault-free; drop
one witness event; jam one platform into silence). The fault-free \fab\ run defines the
ground-truth oracle of mission incidents, $|I^\star|=3$ (collective package, EMCON
breach, unauthorised entry).

\subsection{Detection, Provenance, and Correctness under Fault}
\paragraph{RQ1 / RQ3, detection and provenance.} On the attack, all four per-platform
monitors report \textsf{no\_violation} (each platform transmits $\le 300$\,B, below its
local limit, on an authorised sensor). The compositional monitor detects all three
mission incidents and returns provenance naming the contributing platforms and witness
events, e.g. the package incident with agents \{\texttt{uav\_1..ugv\_1}\} and witnesses
\{\texttt{uav\_1-0,\dots,ugv\_1-0}\}. Per-platform guardrails detect $0/3$; the collective
violation is visible only through composition.

\paragraph{RQ2 / RQ5, correctness and false all-clears under fault.}
Table~\ref{tab:main} reports, per configuration and fault, how many oracle incidents
are \emph{preserved} (violation), \emph{downgraded} to a flagged unknown, or
\emph{silently missed} as a false all-clear. The central best-effort monitor detects
all three fault-free but, under loss and jamming, emits 3 silent false
all-clears in total: a dropped imagery witness makes the package look absent, and a
jammed platform makes both the package and the emissions budget look satisfied. The
\fab\ emits 0: each unresolved incident becomes an explicit
\textsf{unknown/incomplete} (event-id gap) or \textsf{unavailable} (mission-clock
silence), never a clean all-clear.

\begin{table}[t]
\centering\small
\caption{Attack-scenario mission-incident preservation under the fault campaign
($|I^\star|=3$). Cells show preserved incidents; ``$+u$'' = incidents downgraded to a
flagged \textsf{unknown}. The last column is the total number of \emph{silent false
all-clears} across the campaign, the quantity a mission commander must never see. The
local-only (per-platform) baseline has no representation of cross-agent incidents, so it is
treated as producing a mission-level all-clear whenever all per-platform monitors report
compliance.}
\label{tab:main}
\setlength{\tabcolsep}{4pt}
\begin{tabular}{@{}lcccc@{}}
\toprule
\textbf{Configuration} & \textbf{fault-} & \textbf{drop} &
\textbf{jam} & \textbf{silent false} \\
 & \textbf{free} & \textbf{witness} & \textbf{platform} & \textbf{all-clears} \\
\midrule
per-platform guardrails   & $0/3$ & $0/3$ & $0/3$ & $9$ \\
central (best-effort)  & $3/3$ & $2/3$ & $1/3$ & $3$ \\
\fab\ (evidence-aware) & $3/3$ & $2/3\,{+}1u$ & $1/3\,{+}2u$ & $0$ \\
\bottomrule
\end{tabular}
\end{table}

\paragraph{Benign mission.} On the cooperative benign mission (permitted
single-fragment recon, authorised entry, emissions within budget) all three
configurations raise 0 mission-level and 0 local false alarms.

\subsection{Generalisation across Objectives and Tiers}
\begin{figure}[t]
\centering
\begin{tikzpicture}
\begin{axis}[ybar, width=0.72\textwidth, height=4.8cm, ymin=0, ymax=21,
  bar width=26pt, ylabel={silent false all-clears}, ylabel style={font=\small},
  symbolic x coords={per-platform, central, RV-Fabric}, xtick=data,
  nodes near coords, nodes near coords align={vertical},
  every node near coord/.append style={anchor=south, yshift=2pt, font=\small},
  clip=false, tick label style={font=\small}, enlarge x limits=0.85, axis lines=left,
  ymajorgrids, grid style={black!10}, ytick={0,5,10,15}]
\addplot[ybar,bar shift=0pt,fill=black!45] coordinates {(per-platform,18)};
\addplot[ybar,bar shift=0pt,fill=black!25] coordinates {(central,9)};
\addplot[ybar,bar shift=0pt,fill=black!8,draw=black] coordinates {(RV-Fabric,0)};
\end{axis}
\end{tikzpicture}
\caption{Silent false all-clears summed over the four objectives and the three-fault
campaign (lower is better; shading darkens with severity). Per-platform guardrails are
structurally blind (they never see the cross-agent incident); the central best-effort
monitor silently misses under fault; the \fab\ alone reaches zero.}
\label{fig:fac}
\end{figure}
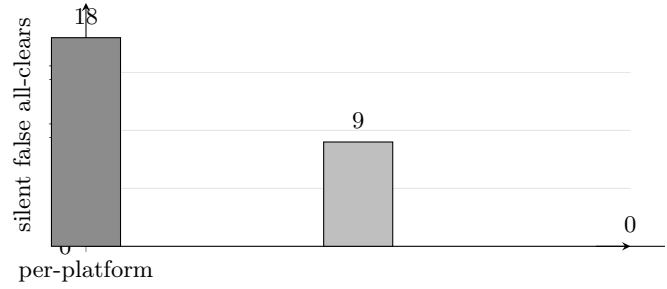

\paragraph{Generalisation across objectives (RQ1/RQ5).} The core result is not specific
to the intelligence-package objective. Table~\ref{tab:obj} repeats the experiment for
four distinct mission objectives spanning four property kinds, a semantic composition
(prohibited package), a swarm-wide sum (collective exfiltration budget), a cross-agent
count (airspace deconfliction), and a cross-agent causal chain (prohibited targeting
workflow). In every case the attack is locally compliant (per-platform monitors detect
$0$), the compositional monitor detects it fault-free, and across the three-fault
campaign the central best-effort monitor accrues 9 silent false all-clears in
total while the \fab\ accrues 0 (Fig.~\ref{fig:fac}). A scale sweep that replicates
the package pattern across squads confirms detection and zero false all-clears at
$4/16/40/100$ platforms, with sub-millisecond mission-monitor evaluation time.

\begin{table}[t]
\centering\small
\caption{Silent false all-clears summed over the three-fault campaign, per objective and
configuration. Every objective: per-platform monitors are blind, the central monitor
silently misses under fault, and the \fab\ never does.}
\label{tab:obj}
\setlength{\tabcolsep}{4pt}
\resizebox{\textwidth}{!}{%
\begin{tabular}{@{}R{4.1cm}R{2.9cm}ccc@{}}
\toprule
\textbf{Mission objective} & \textbf{Property kind} & \textbf{per-platform} & \textbf{central} & \textbf{\fab} \\
\midrule
prohibited intel package ($|I^\star|{=}3$) & semantic composition & $9$ & $3$ & $0$ \\
collective exfiltration budget & swarm-wide sum & $3$ & $2$ & $0$ \\
airspace deconfliction & cross-agent count & $3$ & $2$ & $0$ \\
prohibited targeting workflow & cross-agent causal chain & $3$ & $2$ & $0$ \\
\midrule
\textbf{total} & & $18$ & $9$ & $0$ \\
\bottomrule
\end{tabular}}
\end{table}

The three additional objectives exercise the same structure through different property
kinds. \emph{Collective exfiltration}: each platform transmits within its local limit but
the swarm total exceeds $B_{\text{mission}}$ (a sum no platform can compute), and jamming
one makes the remaining total look compliant. \emph{Airspace deconfliction}: three UAVs
each legitimately enter a sensitive zone but the joint occupancy exceeds $k$ (a cross-agent
count), and a lost entry hides it unless the gap is observed. \emph{Prohibited targeting
workflow}: platforms split a detect~$\prec$~classify~$\prec$~geolocate~$\prec$~share
sequence so no platform does more than one step (a cross-agent causal chain), and a
missing step downgrades to \textsf{unknown}. That the same composition yields zero silent
false all-clears across a semantic composition, a sum, a count, and a causal chain shows
the guarantee is a property of the framework, not one hand-picked scenario.

\paragraph{Genuine three-tier composition (L1$\to$L2$\to$L3).} To exercise the squad tier as
more than a pass-through, we split the four package fragments across \emph{two} squads of
three platforms each: Squad 1 (\texttt{uav\_1} coords, \texttt{uav\_2} imagery, \texttt{uav\_3}
comms) and Squad 2 (\texttt{ugv\_1} access-route, plus a benign relay and a benign recon
platform). Each squad coordinator (L2) aggregates only its own platforms into a \emph{partial}
profile, and the mission monitor (L3) composes the two partials (Table~\ref{tab:hier}).
Fault-free, Squad 1 reports \texttt{partial\{coords,imagery,comms\}} and Squad 2
\texttt{partial\{access\}}; neither partial is a violation, yet their L3 union is the
complete prohibited profile (\textsf{violation}, with provenance spanning both squads).
Crucially, completeness propagates \emph{up} the hierarchy: dropping \texttt{uav\_2}'s imagery
witness makes Squad 1's partial \textsf{incomplete}, the evidence-aware L3 inherits that tag and
returns \textsf{unknown}, whereas a best-effort L3 that ignores squad completeness silently
clears the mission. Platform compliance (L1), squad partial-aggregation (L2), and mission
composition (L3) are thus each load-bearing.

\begin{table}[t]\centering\small
\caption{Two-squad hierarchical composition. Each squad's L2 emits only a \emph{partial}
profile; only the L3 composition of both partials is the prohibited profile. Under a dropped
in-squad witness, squad completeness propagates to the L3 verdict.}
\label{tab:hier}
\setlength{\tabcolsep}{4pt}
\begin{tabular}{@{}R{2.9cm}R{3.7cm}R{2.4cm}R{2.0cm}@{}}
\toprule
\textbf{Scenario} & \textbf{L2[Sq1]} & \textbf{L2[Sq2]} & \textbf{L3 verdict} \\
\midrule
fault-free (fabric)    & \{coords,imagery,comms\}, sound      & \{access\}, sound & violation \\
drop imagery (fabric)  & \{coords,comms\}, \emph{incomplete}  & \{access\}, sound & unknown \\
drop imagery (central) & \{coords,comms\}, sound              & \{access\}, sound & no\_violation$^\dagger$ \\
\bottomrule
\end{tabular}\\[2pt]
{\footnotesize $^\dagger$ silent false all-clear: the best-effort L3 cannot see the in-squad loss.}
\end{table}

\subsection{Robustness, Attribution, and Sensitivity}
\paragraph{The split is produced by a real LLM planner.} The task-split attack is
not scripted: given the mission order and the injected intel note, a real LLM planner emits
the per-platform actions that compose the violation. A separate empirical study
drives this attack through seven planners and several delivery variants (subtle, directive,
per-platform \emph{distributed} injection, benign-domain reframing, and authorisation forgery),
finding susceptibility to be model-dependent and non-monotonic (weaker open models comply, the
strongest hosted models refuse the overt tasking yet comply once it is reframed or delivered one
innocuous fragment at a time), while the compositional monitor detects $100\%$ of the splits that
are \emph{produced} (by construction, once the witness events are observed). Here we take the attack
as given and focus on evidence-aware \emph{detection} and composition under fault.

\paragraph{Provenance, robustness, and causal attribution.} Three further experiments
sharpen the picture. \emph{Provenance (RQ3):} for every detected violation across the four
objectives, the reported platform set exactly matches the ground-truth contributing
platforms: across the six detected violation instances for which a ground-truth
contributing set is defined, the reported provenance set matched exactly in all six
($6/6 = 100\%$), so a supervisor always learns
\emph{which} platforms jointly breached the policy. \emph{Robustness:} replaying the
\emph{benign} mission under the same faults produces 0 false violations (only
\textsf{unknown}s): evidence loss can withhold a verdict but never fabricate one, so the
verification-aware fabric adds no false alarms. \emph{Causal attribution:} ablating individual
guarantees (Table~\ref{tab:abl}) shows each fabric signal prevents a distinct class of
silent miss, removing the event-id gap check (G1) reintroduces the silent miss under
\emph{loss}, and removing the mission clock (G3) reintroduces it under \emph{jamming}, so the zero-false-all-clear result is causally attributable to the fabric, not to the
scenario. Finally, the mission monitor evaluates a $1000$-platform swarm in $1.2$\,ms,
so mission-level checking is far below the second-scale mission-decision budget.
\emph{Statistical stability:} over $500$ randomized task-split missions (random platform
pool, random benign background and timing) the fabric detects every planted violation
and, under a randomly dropped witness, silently clears a real violation in $0\%$ of
missions versus $100\%$ for the central monitor, the result is a property of the
design, not of a particular mission instance.

\paragraph{Threat-model sensitivity.} Detection is not brittle to the adversary's
parameters. Table~\ref{tab:squeeze} sweeps the \emph{split cardinality} of the
collective-exfiltration objective (swarm total fixed, per-platform local limit $500$\,B,
mission EMCON budget $900$\,B): a small split pushes each platform's emission over its
local limit, so the per-platform (L1) monitor catches it; a large split stays locally
compliant, but the swarm-wide sum trips the cross-agent (L3) EMCON property. No split
cardinality evades both tiers at once, so the adversary is squeezed between the
per-platform limit and the cross-agent threshold. Beyond loss and jamming, the fault
matrix also covers \emph{reordering}, where the fabric's trusted ingest order prevents the
false alarm a naive arrival-order monitor raises, and a \emph{slow consumer}, where
independent cursors preserve full detection (zero false all-clears) at bounded latency.

\begin{table}[t]\centering\small
\caption{Threat-model sensitivity: split cardinality of the collective-exfiltration
objective (swarm total $1200$\,B, per-platform local limit $500$\,B, EMCON budget
$900$\,B). The adversary is squeezed: too few platforms trip the local limit, too many are
caught by the cross-agent sum.}
\label{tab:squeeze}
\setlength{\tabcolsep}{6pt}
\begin{tabular}{@{}ccccl@{}}
\toprule
\textbf{split $N$} & \textbf{B/platform} & \textbf{L1 (local)} & \textbf{L3 (EMCON)} & \textbf{caught by} \\
\midrule
$2$ & $600$ & violation & violation & L1 (local) \\
$3$ & $400$ & clean     & violation & L3 (cross-agent) \\
$4$ & $300$ & clean     & violation & L3 (cross-agent) \\
$6$ & $200$ & clean     & violation & L3 (cross-agent) \\
\bottomrule
\end{tabular}
\end{table}

\begin{table}[t]
\centering\small
\caption{Guarantee ablation on the intel-package objective: silent false all-clears when a
single fabric signal is removed. G1's event-id gap detects the dropped witness; G3's
mission clock detects the jammed platform; each removal reintroduces exactly one class of
silent miss.}
\label{tab:abl}
\setlength{\tabcolsep}{6pt}
\begin{tabular}{@{}lcc@{}}
\toprule
\textbf{Configuration} & \textbf{drop-witness (loss)} & \textbf{jam-platform (silence)} \\
\midrule
full fabric              & $0$ & $0$ \\
$-$G1 (no gap detection) & $1$ & $0$ \\
$-$G3 (no mission clock) & $0$ & $2$ \\
\bottomrule
\end{tabular}
\end{table}

\paragraph{Timing, skew, and benign-collaboration sensitivity.} Three sweeps probe robustness
to timing parameters and to legitimate collaboration. \emph{Silence timeout:} the mission-clock
silence rule trades jam-detection latency against false-silence flags; over $300$ jittery
missions with heavy-tailed comms stalls, a $2$\,s timeout raises $\approx2.9$ false-silence
flags per mission, a $5$\,s timeout $\approx0.5$ (jam detected in $\approx4$\,s), and $10$\,s
zero (but $\approx9$\,s latency): $5$\,s is the knee. \emph{Clock skew:} applying a per-platform
clock offset to the order-dependent monitors, the verdict is preserved below the order window and
\emph{downgrades to \textsf{unknown}} beyond it (authorisation-order at skew $\ge$$10$\,s for a
$5$\,s window; targeting-chain once the observed step order inverts) rather than emitting a false
verdict, while the order-insensitive occupancy count is unaffected at every skew; this motivated
extending the chain monitor with an explicit order-violation$\to$\textsf{unknown} trigger.
\emph{Benign collaboration:} across seven legitimate scenarios that superficially resemble the
attacks (authorised full-profile collection, the same fragments on an unprotected target,
aborted (three-of-four) collection, authorised emergency entry, an authorised targeting workflow,
sub-$k$ occupancy, and within-budget emissions), the monitors raise 0 false alarms,
confirming the composition is correctly conditioned on target, authorisation, completion, and
count (and prompting the same authorisation conjunct to be added to the targeting-workflow
property).

\subsection{Systems, Formal-Engine, and Advanced-Fault Validation}
\paragraph{Systems validation on the reused RV-Fabric.} We validate the L2 transport guarantees
by executing them \emph{live on the evaluation host} against the reused RV-Fabric
implementation: real Mosquitto\,$+$\,NATS JetStream with the gateway durable
outbox recast as the L2 coordinator. These runs exercise the fabric and the L3 mission monitor;
integrating the ROS\,2/DDS front hop into a single end-to-end path is future work.
\emph{End-to-end latency} over the gen$\to$MQTT$\to$L2$\to$JetStream$\to$L3 path ($n{=}1000$ live
mission events) has a fault-free median of $3.7$\,ms and, under a degraded link ($10\%$ drop with
durable retransmit), a $42$\,ms median with no accepted event lost ($1000/1000$
delivered), comfortably within the second-scale mission-decision budget. \emph{Coordinator
crash/recovery}: once the outbox entry is persisted, 0 accepted verdicts are lost and the
after-publish replay yields 0 duplicate incidents after event-id de-duplication.
\emph{Overload/retention}: \texttt{discard=old} silently drops the oldest verdicts under sustained
overload, whereas \texttt{discard=new} surfaces overflow as an explicit health alert \emph{before}
continuity is lost. \emph{Consumer isolation}: independent durable cursors let a fast L3 consumer
drain its backlog while a slow peer lags, so one lagging monitor cannot stall the mission verdict.
These confirm the guarantees hold on the real fabric, not only in the
in-process emulation used for the controlled campaign.

\paragraph{Real formal engine.} The cross-agent primitive is not only decided by our
reference monitor: we export the coordinated-collection property, $\ge 3$ distinct
platforms collect on the protected site within $30$\,s, the swarm analogue of the
edge-IoT cross-gateway property, to the real MonPoly first-order
past-time engine and evaluate it on the attack trace. MonPoly reports the property
satisfied at the third and fourth contributing platform (two satisfaction points),
confirming that the distributed violation is detectable by a genuine formal engine
carried over the fabric, not by a bespoke checker.

\paragraph{Adaptive adversary, integrity faults, and detection latency.} Four
experiments probe dimensions beyond loss and jamming. \emph{Adaptive (low-and-slow)
adversary:} a patient adversary spreads the four package parts by an inter-part gap $S$,
giving a collection span $3S$; a windowed monitor detects the package iff its window
$W \ge 3S$. Because the \fab\ aggregates over the whole mission window, evasion requires
spreading collection \emph{across} missions, a documented limit
(Sect.~\ref{sec:threats}), at a buffer-state cost proportional to $W$. \emph{Detection
latency:} under online, streaming evaluation the fault-free verdict fires at the first event
that completes the package, with no additional logical waiting interval beyond witness
completion (the per-event processing cost is microseconds, reported separately); under a
dropped witness the \fab\ raises a
\emph{finite}-latency \textsf{unknown} the instant the event-id gap becomes observable,
whereas the central monitor's latency is \emph{unbounded}: it never signals (the silent
miss made quantitative). \emph{Byzantine (lying) platform (exploratory extension beyond the
core threat model of Sect.~\ref{sec:threat}):} when each platform \emph{under-reports} its
own emissions (self-reported swarm total $800$\,B, below the $900$\,B budget), a
self-report-only monitor is fooled; an independent RF-sentry witness observes the true
$1200$\,B and the corroboration property fires, with integrity provenance naming the
under-reporting platforms. Jamming the sentry yields \textsf{unknown}, not a silent clear:
an integrity fault is thus handled like a loss fault \emph{provided} an independent
corroborating witness exists, which is the assumption this extension adds. \emph{Recovery:} a
withheld \textsf{unknown} resolves to \textsf{violation} once a dark contributor returns, so
the downgrade is provisional and sound; the central monitor's blackout all-clear, by
contrast, has already been delivered.

\paragraph{No silent all-clears, and graceful degradation.} We collapse the entire fault
campaign into a single \emph{No-Silent-Clear Rate},
\[
  \text{NSCR}=1-\frac{\text{silent false all-clears}}{|I^\star_{\text{campaign}}|},
\]
the fraction of oracle incidents that are \emph{not} silently cleared, i.e.\ reported as
\textsf{violation} or honestly downgraded to \textsf{unknown} (Table~\ref{tab:advanced}). We
prefer this to a ``verdict-preservation'' count because an honest \textsf{unknown} does not
preserve the original \textsf{violation}; it merely refuses to clear it, which is exactly
the safety property we claim. The \fab\ scores $100\%$, the central monitor $50\%$, and
per-platform guardrails $0\%$; the strict violation-vs-\textsf{unknown} split is the ``$+u$''
breakdown already shown in Table~\ref{tab:main}. A compound-fault sweep confirms the
guarantee is not fragile to fault \emph{count}: as up to eight squads are simultaneously
faulted, the \fab's silent-clear count stays $0$ and its \textsf{unknown} count
rises monotonically (graceful degradation), while the central monitor's silent misses climb
to $8$.

\paragraph{Is the property a tuned knob? (ROC).} To check that the collective-budget
property is well-posed rather than a hand-fitted threshold, we generate $400$ missions whose
true swarm-emission total is drawn uniformly over $[500,1300]$\,B (a mission is a
\emph{positive} iff its true total exceeds the $900$\,B reference budget), split the total
across four platforms, and add independent Gaussian measurement noise ($\sigma=60$\,B) per
platform. Sweeping the decision threshold and scoring each mission by its \emph{summed}
measured emissions gives area-under-ROC $0.96$ (near-perfect separation, so detection is not
a tuned artefact); scoring instead by any \emph{single} platform's measurement gives $0.50$
(chance), because a per-platform view structurally cannot observe a swarm-wide sum. The
otherwise-deterministic property admits an ROC at all only because the injected
\emph{measurement noise}, not the logic, makes the thresholded sum a non-trivial classifier.
We stress that ROC/AUC is \emph{not} the headline instrument: a strict detection ROC treats
an honest \textsf{unknown} and a silent miss identically (both are non-detections), so it
cannot credit the evidence-aware behaviour that is the whole point, which is exactly why
NSCR, not AUC, is the metric of record. Finally, when all six properties fire on a single
mission trace, provenance disentangles which platforms triggered which property, and the
command node that issued the authorisation appears in \emph{no} violation (no collateral
blame).

\begin{table}[t]
\centering\small
\caption{No-Silent-Clear Rate (fraction of oracle incidents \emph{not} silently cleared,
i.e.\ reported as a violation or honestly downgraded to \textsf{unknown}) over the full fault
campaign, and silent false all-clears when eight squads are faulted simultaneously. The
\fab\ never silently clears, at any fault count.}
\label{tab:advanced}
\setlength{\tabcolsep}{5pt}
\begin{tabular}{@{}lcc@{}}
\toprule
\textbf{Configuration} & \textbf{NSCR} & \textbf{Silent all-clears} \\
                       &               & \textbf{(8 squads faulted)} \\
\midrule
per-platform guardrails & $0\%$   & n/a \\
central (best-effort)   & $50\%$  & $8$ \\
\fab\ (evidence-aware)  & $100\%$ & $0$ \\
\bottomrule
\end{tabular}
\end{table}

\paragraph{Per-mechanism walkthrough.} The three faults isolate distinct failure modes.
Under \emph{drop imagery witness} (a delivery loss, G1), the collective-package property
loses one of its four witnesses and cannot fire. The central monitor, lacking any
completeness signal, reports \textsf{no\_violation} on the package, a \emph{silent
false all-clear} indistinguishable from a genuinely benign mission, while still
detecting the two unaffected incidents ($2/3$, one silent miss). The evidence-aware
fabric instead observes an event-id gap on \texttt{uav\_2}'s stream and downgrades the
package verdict to $(\textsf{unknown},\textsf{incomplete})$: a flagged $2/3\,{+}1u$ with
zero silent misses. Under \emph{jam \texttt{uav\_3}} (a denial, G3), both the package
(missing \texttt{comms}) and the emissions budget (missing \texttt{uav\_3}'s
transmissions, so the swarm total falls below $B_{\text{mission}}$) appear satisfied to
the central monitor, \emph{two} silent false all-clears ($1/3$), whereas the
mission clock surfaces \texttt{uav\_3} as a missing stream and both verdicts become
\textsf{unknown}/\textsf{unavailable} ($1/3\,{+}2u$, zero silent misses). Across the
campaign the central monitor accrues three silent false all-clears and the fabric zero:
the fabric never trades a \textsf{no\_violation} for missing evidence.

\subsection{Cost, Summary, and Assurance Workflow}
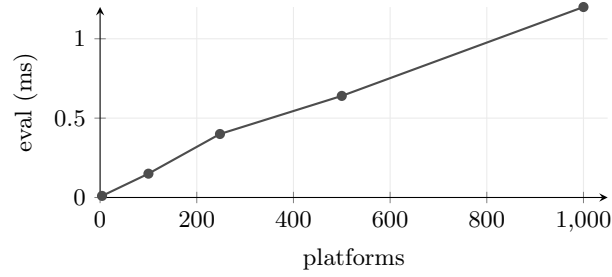
\begin{figure}[t]
\centering
\begin{tikzpicture}
\begin{axis}[width=0.68\textwidth, height=4.1cm, xlabel={platforms}, ylabel={eval (ms)},
  xlabel style={font=\small}, ylabel style={font=\small}, tick label style={font=\small},
  ymin=0, xmin=0, xmax=1050, mark size=1.5pt, axis lines=left,
  grid=major, grid style={black!8}]
\addplot[mark=*,thick,black!70] coordinates {(4,0.01)(100,0.15)(248,0.40)(500,0.64)(1000,1.2)};
\end{axis}
\end{tikzpicture}
\caption{Mission-monitor evaluation time vs swarm size (real measurements). Cross-agent
checking scales roughly linearly and stays sub-$2$\,ms to $1000$ platforms, orders of
magnitude below the second-scale mission-decision budget.}
\label{fig:scale}
\end{figure}

\paragraph{RQ4, cost.} The monitors are past-time and stream-based and add per-event
overhead comparable to the single-agent case (microseconds/event); the mission-level
cost is dominated by the second-scale mission dynamics, and the fabric's
delivery/recovery costs are those measured for the underlying transport
(crash recovery linear in the un-acked backlog; $0\%$ loss to
$1000$ producers). Fig.~\ref{fig:scale} plots the measured mission-monitor evaluation time against swarm
size: cross-agent checking scales roughly linearly and stays under $2$\,ms to $1000$
platforms. Beyond monitor evaluation, end-to-end verdict latency across the full fabric path
($t_{\text{verdict}}-t_{\text{event}}$: event $\to$ MQTT $\to$ squad sidecar $\to$ JetStream
$\to$ L3) is measured \emph{live} on the real Mosquitto\,$+$\,NATS JetStream stack, a
fault-free median of $3.7$\,ms and a degraded-link median of $42$\,ms with no accepted event lost
(reported in full under \emph{Systems validation on the reused RV-Fabric} above), well within
a second-scale mission-decision budget.

\paragraph{Summary.} The central result is precisely the mission-assurance claim: \emph{every
per-platform monitor reports compliance, yet the compositional monitor detects the
mission-level violation and names the platforms that jointly produced it; and under
contested communications the verification-aware fabric withholds the all-clear rather than
silently missing the attack.}

\paragraph{Mission-assurance workflow.} The verdicts are an audit trail, not a terminal
pass/fail. Provenance lets a supervisor, on a \textsf{violation}, see \emph{which}
platforms jointly breached the policy, and on an \textsf{unknown/incomplete} verdict know
that a decision is withheld for want of evidence rather than granted by default, a
distinction that is operationally decisive under jamming: ``we cannot currently certify
Sector B'' lets a commander hold or re-task, whereas a silent all-clear invites
exactly the adversary's intended action. Since the monitors are decoupled from the flight
controllers (Sect.~\ref{sec:threat}), the assurance layer drops onto an existing autonomy
stack without touching safety-critical control.

\subsection{Threats to Validity}
\label{sec:threats}
The quantitative results are \emph{controlled invariant-preservation experiments} (a
deterministic kinematic mission with a known ground-truth oracle, designed to isolate
whether each design element preserves the mission verdict under fault) and not a
population-level measurement of operational detection accuracy; the clean $0/100\%$ figures
should be read in that light. Of the realism layer, the ROS\,2/DDS middleware integration and
the ArduPilot SITL flight-dynamics path are both \emph{executed} (Sect.~\ref{sec:impl},
Fig.~\ref{fig:mission}); only Gazebo \emph{visual} rendering (a GPU concern, cosmetic) and a
larger physical, at-scale sweep remain future work. The semantic-extraction path is deliberately secondary; text-only violations depend on
classifier quality. Core-model platform integrity beyond going silent is out of scope (with
the single exploratory RF-corroboration extension of Sect.~\ref{sec:eval}), as is a full
formal soundness proof of the composition algebra (adapted from the prior edge-IoT fabric).

\section{Related Work}
\label{sec:related}
\paragraph{Agent safety and multi-step attacks.} Single-agent guardrails and policy
monitors check one agent's actions against rules, but sequential tool-attack chains and
prompt-injection benchmarks show that a large fraction of multi-step attacks defeat
per-event checks~\cite{stac2025,agentdojo2024,zhan2024injecagent}; and in multi-agent
settings an injection can even propagate agent-to-agent~\cite{promptinfection}. Our single-agent RV of LLM
actions composes spatial, temporal, and semantic monitors for
\emph{one} agent and establishes that composition is necessary rather than convenient;
the present work lifts that composition to the \emph{swarm}, where the violation is
distributed across platforms and no single-agent view, however well composed, suffices.

\paragraph{Distributed and spatial RV.} Reaching a verdict across faulty, asynchronous
monitors is studied at the algorithm
level~\cite{fraigniaud2014opinions,bonakdarpour2016decentralized}, and spatial and
spatio-temporal logics monitor properties over networks of
components~\cite{strel2017}. That line makes the monitoring \emph{logic} fault- or
reordering-tolerant while abstracting the transport. We are complementary and beneath
it: the verification-aware fabric makes the delivery layer's loss,
ordering, and liveness first-class and observable, so the stream the swarm monitor
consumes, and reasons about with any of those algorithms, is the stream their
correctness assumes. This is what a swarm monitor needs under contested communications,
and it is what lets us report an honest \textsf{unknown} instead of a silent all-clear.
Our own two-level hierarchical monitoring line (edge device to gateway to backend, with
TeSSLa- and MonPoly-based first-order correlation) established these tiers for edge-IoT
security monitoring and cyber-physical
resilience~\cite{ares2026,csr2026,kekatos2026resilagent,lekidis2026ipv6}; the
present paper carries that hierarchy to a cooperative LLM-assisted swarm.

\paragraph{Runtime assurance for robotics.} Runtime-assurance architectures switch to a
verified safety controller when a monitor flags danger, from the Simplex
architecture~\cite{sha2001simplex} to robotics frameworks such as SOTER~\cite{desai2019soter};
and ROS-level monitors such as ROSMonitoring~\cite{ferrando2021rosmonitoring} attach runtime
verification to a robot's message bus; feedback controllers can themselves be synthesised to
satisfy temporal-logic specifications~\cite{dang2023cegil}. These secure a \emph{single}
platform's control and safety envelope. Our concern is orthogonal and one level up: a \emph{mission-level,
cross-agent} obligation that no single platform's monitor can express, checked over an
evidence-aware transport so that contested or intermittent communication yields an explicit
\textsf{unknown} rather than a silent all-clear. The single-platform assurance approaches
reviewed here do not explicitly address compositional multi-robot \emph{mission} obligations
over an adversarial, lossy evidence channel.

\paragraph{M\&S of autonomous systems.} Modelling and simulation is the established
setting for evaluating autonomous-system assurance under repeatable, injected
conditions~\cite{mesas2024}. Digital-twin and FMI co-simulation approaches assure
autonomous systems by running analyses against a high-fidelity
model~\cite{temperekidis2022dt,temperekidis2022fmirv,abdelsalam2025dtanes}, within a broader
continuous-engineering methodology for trustworthy learning-enabled autonomous
systems~\cite{bensalem2023foceta}, and design-time formal analysis can check learning-enabled
components directly~\cite{eleftheriadis2022nnequiv}; these are complementary to this
work and the subject of a companion paper. Our contribution is orthogonal: a
simulation-evaluated, compositional, evidence-aware monitor for the swarm's
\emph{mission-level} obligations, whose verdicts are auditable mission-assurance
evidence rather than a pass/fail flag.

\section{Conclusion}
\label{sec:conclusion}
We presented a three-tier, evidence-aware compositional runtime-verification framework
for LLM-assisted autonomous robot swarms on ISR missions in contested environments. A
distributed prompt-injection attack that is compliant at every platform is invisible to
per-platform monitors yet detected, with platform-level provenance, by the compositional
monitor; and under loss and jamming the verification-aware fabric emits zero silent
false all-clears where a best-effort central monitor emits several. The framework lifts
hierarchical agent-safety monitoring from the single agent to the swarm and turns
transport-level guarantees into auditable, platform-attributed mission-assurance
evidence. Future work adds Gazebo visual rendering and a larger physical, at-scale sweep over
the executed ROS\,2/ArduPilot layer, and a formal soundness argument for the composition
algebra.

\bibliographystyle{splncs04}
\bibliography{references}

@inproceedings{ares2026,
  author    = {Kekatos, Nikolaos and Chintri, Marinelio and Katsaros, Panagiotis and
               Lekidis, Alexios and Nianios, Tom and Seitoglou, Ioannis and
               Temperekidis, Anastasios and Basagiannis, Stylianos},
  title     = {A Hierarchical Runtime-Verification Approach for Security Monitoring of Edge-IoT},
  booktitle = {Int. Workshop on Advances in Practical Security (ARES)},
  year      = {2026}
}

@inproceedings{csr2026,
  author    = {Kekatos, Nikolaos and Chintri, M. and Katsaros, Panagiotis and
               Nianios, Tom and Seitoglou, I. and Basagiannis, Stylianos},
  title     = {Hierarchical Security Monitoring for Edge-IoT: A Formal Methods Approach},
  booktitle = {IEEE International Conference on Cyber Security and Resilience (CSR)},
  year      = {2026}
}

@inproceedings{eleftheriadis2022nnequiv,
  author    = {Eleftheriadis, Charis and Kekatos, Nikolaos and Katsaros, Panagiotis and Tripakis, Stavros},
  title     = {On Neural Network Equivalence Checking Using {SMT} Solvers},
  booktitle = {Formal Modeling and Analysis of Timed Systems (FORMATS)},
  series    = {LNCS}, pages = {237--257}, year = {2022}
}

@article{abdelsalam2025dtanes,
  author    = {Abdelsalam, Mohamed and Bensalem, Saddek and Delacourt, Antoine and He, Weicheng and
               Katsaros, Panagiotis and Kekatos, Nikolaos and Ruiz Nolasco, Ricardo and Peled, Doron and
               Ponchant, Matthieu and Ryad, Ismail and Temperekidis, Anastasios and Wu, Changshun},
  title     = {Digital Twin for the Formal Analysis of a Depth of Anesthesia Controller},
  journal   = {SIMULATION}, volume = {101}, number = {3}, pages = {341--360}, year = {2025}
}

@inproceedings{dang2023cegil,
  author    = {Dang, Thao and Donz\'e, Alexandre and Haque, Inzemamul and Kekatos, Nikolaos and Saha, Indranil},
  title     = {Counter-Example Guided Imitation Learning of Feedback Controllers from Temporal Logic Specifications},
  booktitle = {IEEE Conference on Decision and Control (CDC)},
  year      = {2023}
}

@inproceedings{bensalem2023foceta,
  author    = {Bensalem, Saddek and Katsaros, Panagiotis and Kekatos, Nikolaos and others},
  title     = {Continuous Engineering for Trustworthy Learning-Enabled Autonomous Systems},
  booktitle = {Bridging the Gap Between AI and Reality (AISoLA 2023)},
  series    = {LNCS}, volume = {14380}, publisher = {Springer}, year = {2023}
}

@inproceedings{kekatos2026resilagent,
  author    = {Kekatos, Nikolaos and Koutidis, Georgios and Antonakopoulos, Konstantinos and
               Basagiannis, Stylianos and Katsikas, Sokratis and Kavallieratos, Georgios and
               Lekidis, Alexios and Nianios, Tom and Papageorgiou, Elpiniki},
  title     = {{RESILAGENT}: A Three-Pillar Architecture for Cyber-Physical Resilience in Electrical Power and Energy Systems},
  booktitle = {ARES 2026 Workshops (EPES-SPR)},
  year      = {2026}
}

@inproceedings{lekidis2026ipv6,
  author    = {Sinha, Priyanka and Kekatos, Nikolaos and Basagiannis, Stylianos and
               Bruto da Costa, Antonio Anastasio and Lekidis, Alexios and Mitra, Pabitra and
               Nianios, Tom and Papageorgiou, Elpiniki},
  title     = {Explainable Rule Mining of {IPv6} Extension-Header Presence Patterns from Paired-Vantage Captures},
  booktitle = {ARES 2026 Workshops (GenXSec)},
  year      = {2026}
}

@article{basin2015mfotl,
  author = {Basin, David and Klaedtke, Felix and M{\"u}ller, Samuel and Z{\u{a}}linescu, Eugen},
  title = {Monitoring Metric First-Order Temporal Properties},
  journal = {Journal of the ACM}, volume = {62}, number = {2}, year = {2015}
}

@inproceedings{monpoly,
  author = {Basin, David and Klaedtke, Felix and Zalinescu, Eugen},
  title = {The {MonPoly} Monitoring Tool},
  booktitle = {RV-CuBES}, year = {2017}
}

@inproceedings{rtlola,
  author = {Faymonville, Peter and Finkbeiner, Bernd and Schledjewski, Malte and Schwenger, Maximilian and Stenger, Marvin and Tentrup, Leander and Torfah, Hazem},
  title = {{StreamLAB}: Stream-based Monitoring of Cyber-Physical Systems},
  booktitle = {Computer Aided Verification (CAV)}, year = {2019}
}

@inproceedings{dejavu,
  author = {Havelund, Klaus and Peled, Doron and Ulus, Dogan},
  title = {First-Order Temporal Logic Monitoring with {BDD}s},
  booktitle = {Formal Methods in Computer-Aided Design (FMCAD)}, year = {2017}
}

@inproceedings{tessla,
  author = {Convent, Lukas and Hungerecker, Sebastian and Leucker, Martin and Scheffel, Torben and Schmitz, Malte and Thoma, Daniel},
  title = {{TeSSLa}: Temporal Stream-based Specification Language},
  booktitle = {SBMF}, year = {2018}
}

@article{bartocci2018introduction,
  author = {Bartocci, Ezio and Falcone, Yli{\`e}s and Francalanza, Adrian and Reger, Giles},
  title = {Introduction to Runtime Verification},
  journal = {Lectures on Runtime Verification, LNCS 10457}, year = {2018}
}

@inproceedings{bonakdarpour2016decentralized,
  author = {Bonakdarpour, Borzoo and Fraigniaud, Pierre and Rajsbaum, Sergio and Rosenblueth, David A. and Travers, Corentin},
  title = {Decentralized Asynchronous Crash-Resilient Runtime Verification},
  booktitle = {CONCUR}, year = {2016}
}

@inproceedings{fraigniaud2014opinions,
  author = {Fraigniaud, Pierre and Rajsbaum, Sergio and Travers, Corentin},
  title = {On the Number of Opinions Needed for Fault-Tolerant Run-Time Monitoring in Distributed Systems},
  booktitle = {Runtime Verification (RV)}, year = {2014}
}

@inproceedings{strel2017,
  author = {Bartocci, Ezio and Bortolussi, Luca and Loreti, Michele and Nenzi, Laura},
  title = {Monitoring Mobile and Spatially Distributed Cyber-Physical Systems ({STREL})},
  booktitle = {MEMOCODE}, pages = {146--155}, year = {2017}
}

@article{temperekidis2022fmirv,
  author = {Temperekidis, Anastasios and Kekatos, Nikolaos and Katsaros, Panagiotis},
  title = {Runtime Verification for {FMI}-based Co-simulation},
  journal = {Runtime Verification (RV), LNCS 13498}, year = {2022}
}

@inproceedings{temperekidis2022dt,
  author = {Temperekidis, Anastasios and Kekatos, Nikolaos and Katsaros, Panagiotis and others},
  title = {Towards a Digital Twin Architecture with Formal Analysis Capabilities for Learning-Enabled Autonomous Systems},
  booktitle = {MESAS}, series = {LNCS}, volume = {13866}, year = {2022}
}

@misc{stac2025,
  author = {Li, Jing-Jing and He, Jianfeng and Shang, Chao and Kulshreshtha, Devang and
            Xian, Xun and Zhang, Yi and Su, Hang and Swamy, Sandesh and Qi, Yanjun},
  title = {{STAC}: When Innocent Tools Form Dangerous Chains to Jailbreak {LLM} Agents},
  year = {2025}, note = {arXiv:2509.25624}
}

@inproceedings{agentdojo2024,
  author = {Debenedetti, Edoardo and others},
  title = {{AgentDojo}: A Dynamic Environment to Evaluate Prompt Injection Attacks and Defenses for LLM Agents},
  booktitle = {NeurIPS Datasets and Benchmarks}, year = {2024}
}

@article{mqtt,
  author = {{OASIS}}, title = {{MQTT} Version 5.0}, journal = {OASIS Standard}, year = {2019}
}

@misc{nats, author = {{Synadia / CNCF}}, title = {{NATS} and {JetStream}}, year = {2024}, howpublished = {\url{https://nats.io}} }

@article{chung2018aerial,
  author = {Chung, Soon-Jo and Paranjape, Aditya A. and Dames, Philip and Shen, Shaojie and Kumar, Vijay},
  title = {A Survey on Aerial Swarm Robotics},
  journal = {IEEE Transactions on Robotics}, volume = {34}, number = {4}, pages = {837--855}, year = {2018}
}

@article{brambilla2013swarm,
  author = {Brambilla, Manuele and Ferrante, Eliseo and Birattari, Mauro and Dorigo, Marco},
  title = {Swarm Robotics: A Review from the Swarm Engineering Perspective},
  journal = {Swarm Intelligence}, volume = {7}, number = {1}, pages = {1--41}, year = {2013}
}

@inproceedings{ahn2022saycan,
  author = {Ahn, Michael and Brohan, Anthony and Brown, Noah and others},
  title = {Do As I Can, Not As I Say: Grounding Language in Robotic Affordances},
  booktitle = {Conf. on Robot Learning (CoRL)}, year = {2022}
}

@inproceedings{greshake2023injection,
  author = {Greshake, Kai and Abdelnabi, Sahar and Mishra, Shailesh and Endres, Christoph and Holz, Thorsten and Fritz, Mario},
  title = {Not What You've Signed Up For: Compromising Real-World {LLM}-Integrated Applications with Indirect Prompt Injection},
  booktitle = {ACM Workshop on Artificial Intelligence and Security (AISec)}, year = {2023}
}

@article{shi2016edge,
  author = {Shi, Weisong and Cao, Jie and Zhang, Quan and Li, Youhuizi and Xu, Lanyu},
  title = {Edge Computing: Vision and Challenges},
  journal = {IEEE Internet of Things Journal}, volume = {3}, number = {5}, year = {2016}
}

@proceedings{mesas2024,
  editor = {Mazal, Jan and Fagiolini, Adriano and others},
  title = {Modelling and Simulation for Autonomous Systems (MESAS 2024), Revised Selected Papers},
  series = {LNCS}, volume = {15761}, publisher = {Springer}, year = {2025}
}

@article{sha2001simplex,
  title={Using Simplicity to Control Complexity},
  author={Sha, Lui},
  journal={IEEE Software},
  volume={18}, number={4}, pages={20--28}, year={2001}, publisher={IEEE}
}

@inproceedings{desai2019soter,
  title={{SOTER}: A Runtime Assurance Framework for Programming Safe Robotics Systems},
  author={Desai, Ankush and Ghosh, Shromona and Seshia, Sanjit A. and Shankar, Natarajan and Tiwari, Ashish},
  booktitle={49th IEEE/IFIP Int. Conf. on Dependable Systems and Networks (DSN)},
  pages={138--150}, year={2019}
}

@inproceedings{ferrando2021rosmonitoring,
  title={{ROSMonitoring}: A Runtime Verification Framework for {ROS}},
  author={Ferrando, Angelo and Cardoso, Rafael C. and Fisher, Michael and Ancona, Davide and Franceschini, Luca and Mascardi, Viviana},
  booktitle={Towards Autonomous Robotic Systems (TAROS)},
  series={LNCS}, volume={12228}, pages={387--399}, year={2020}, publisher={Springer}
}

@inproceedings{perez2022ignore,
  author = {Perez, Fábio and Ribeiro, Ian},
  title = {Ignore Previous Prompt: Attack Techniques for Language Models},
  booktitle = {NeurIPS ML Safety Workshop}, year = {2022}
}

@inproceedings{liu2024formalizing,
  author = {Liu, Yupei and Jia, Yuqi and Geng, Runpeng and Jia, Jinyuan and Gong, Neil Zhenqiang},
  title = {Formalizing and Benchmarking Prompt Injection Attacks and Defenses},
  booktitle = {USENIX Security Symposium}, year = {2024}
}

@inproceedings{zhan2024injecagent,
  author = {Zhan, Qiusi and Liang, Zhixiang and Ying, Zifan and Kang, Daniel},
  title = {{InjecAgent}: Benchmarking Indirect Prompt Injections in Tool-Integrated Large Language Model Agents},
  booktitle = {Findings of the Association for Computational Linguistics (ACL)}, year = {2024}
}

@misc{promptinfection,
  author = {Lee, Donghyun and Tiwari, Mo},
  title = {Prompt Infection: {LLM}-to-{LLM} Prompt Injection within Multi-Agent Systems},
  year = {2024}, note = {arXiv:2410.07283}
}

@article{leucker2009brief,
  author = {Leucker, Martin and Schallhart, Christian},
  title = {A Brief Account of Runtime Verification},
  journal = {Journal of Logic and Algebraic Programming},
  volume = {78}, number = {5}, pages = {293--303}, year = {2009}
}

@article{bauer2011runtime,
  author = {Bauer, Andreas and Leucker, Martin and Schallhart, Christian},
  title = {Runtime Verification for {LTL} and {TLTL}},
  journal = {ACM Transactions on Software Engineering and Methodology (TOSEM)},
  volume = {20}, number = {4}, year = {2011}
}

\appendix
\section{Reproducibility}
\label{sec:repro}
The reproducible core (kinematic mission, monitors,
verification-aware fabric, fault campaign) and the ROS\,2 adapter are provided as an artifact.
Every \emph{deterministic} quantitative result reproduces on a fresh Ubuntu virtual machine with
\texttt{docker build -t swarm-rv .} then a single \texttt{docker run}, no ROS\,2 or GPU
required: \texttt{run\_demo.py} (Table~\ref{tab:main}); \texttt{experiments.py}
(Table~\ref{tab:obj}, Fig.~\ref{fig:fac}, scale sweep); \texttt{experiments\_extra.py}
(provenance, ablation Table~\ref{tab:abl}, $1000$-platform Fig.~\ref{fig:scale}, and the
$500$ randomized missions at seed~$7$); \texttt{experiments\_advanced.py} (NSCR
Table~\ref{tab:advanced}, compound-fault sweep, and the $400$-mission ROC at seed~$11$);
\texttt{experiments\_hierarchy.py} (two-squad L1$\to$L2$\to$L3, Table~\ref{tab:hier});
\texttt{experiments\_properties.py} (the additional property kinds);
\texttt{threat\_sweep.py} (Table~\ref{tab:squeeze}); and the parameter/robustness sweeps
\texttt{exp\_timeout.py}, \texttt{exp\_skew.py}, and \texttt{exp\_benign.py}. We distinguish two
classes of result. The deterministic RV experiments above are \emph{fully reproducible} (fixed
seeds, no wall-clock or network dependence). The LLM-in-the-loop attack campaign (the seven planners, injection variants, and attack-success rate (ASR)) is reported separately. The systems validation was executed live
against the reused RV-Fabric stack (real Mosquitto\,$+$\,NATS JetStream): end-to-end latency via
\texttt{e2e\_latency.py --real} ($n{=}1000$), and coordinator crash/recovery, overload/retention,
and consumer isolation via \texttt{run\_crash\_exp.sh}, \texttt{exp\_retention.py}, and
\texttt{exp\_isolation.py} from the companion artifact against the live broker.
The ROS\,2 result of
Sect.~\ref{sec:impl} runs via \texttt{ros2\_realism/run\_ros2\_demo.sh} in a
\texttt{ros:humble} container; the ArduPilot SITL mission (Fig.~\ref{fig:mission}) runs via
\texttt{sitl\_mission/run\_sitl\_mission.sh} on an x86 host; and the cross-agent property is
confirmed by the real MonPoly engine via \texttt{swarm\_to\_monpoly.py}. All deterministic RV,
transport, and simulation experiments use fixed seeds and were reproduced on an independent
Ubuntu VM; the LLM outputs are archived together with prompt hashes, model identifiers, request
parameters, and raw responses, but because hosted models may change, exact regeneration of those
outputs is not guaranteed: they are \emph{preserved}, not necessarily reproducible.

\end{document}